\documentstyle[editedbook,epsfig,psfig,epsf]{mq}

\def\cm2{cm$^2$ }
\def\se1{s$^{-1}$ }

\begin{opening}
\title{GX 339--4: back to life}
\author{T. Belloni$^1$, E. Nespoli$^1$, J. Homan$^1$, M. van der Klis$^2$,
	W.H.G. Lewin$^3$,}
\author{J.M. Miller$^3$ \& M. M\'endez$^4$}
\institute{$^1$INAF -- Osservatorio Astronomico di Brera, Merate, Italy. \\
	   $^2$Astronomical Institute ``A.Pannekoek'', 
	   Univ. of Amsterdam, the Netherlands \\
	   $^3$MIT, Cambridge, USA\\
	   $^4$SRON, Utrecht, the Netherlands
	   }
\end{opening}

\runningtitle{GX 339--4 back to life}
\runningauthor{Belloni}

\begin{document}
\vspace{-0.5cm}
\begin{abstract}
{\small 
We report preliminary results of a RossiXTE campaign on 
the 2002 outburst
of the black-hole candidate GX~339--4. We show power density spectra of
five observations during the early phase of the outburst. The first four 
power spectra
show a smooth transition between a Low State and a Very  High State. The
fifth power spectrum resembles a High State, but a strong 6 Hz QPO appears
suddenly within 16 seconds. 
}
\end{abstract}

\section{GX 339--4 as a transient: our campaign}

After almost three years of quiescence, the ``persistent" BHC GX~339--4
became active again on 2002 March 26 \cite{atel}. This source is
important , as in the past 
it has shown in the past all of the ``canonical'' states of
BHCs \cite{mendez}.

\begin{figure}[htb]
\centering
\psfig{file=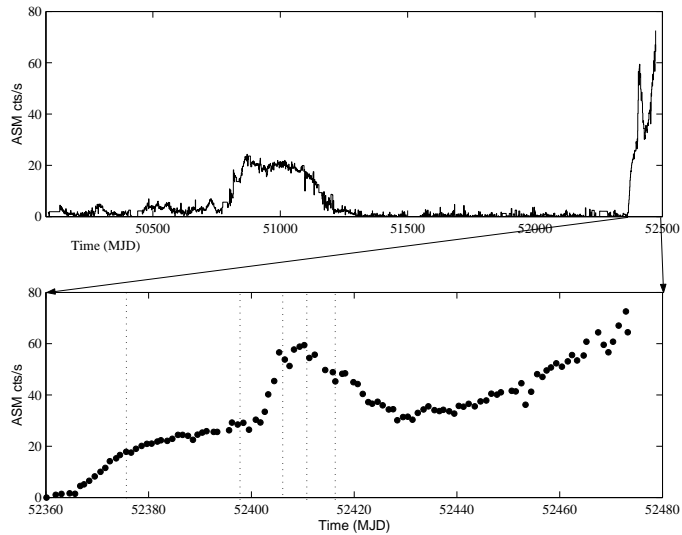,width=9cm}
\caption{Top panel: full ASM light curve for GX~339--4. Bottom panel: zoom
	of the new activity period considered here. The times of our
	pointings are marked by dotted lines.}
\label{fig:ex}
\end{figure}

The full RossiXTE/ASM
light curve from the start of the mission is shown in Fig. 1.
We started our observations
with RossiXTE on April 3rd and obtained roughly one pointing
per week since then. The ASM rate (1.5-1 keV) increased to 0.8 Crab in less
than two months, then decreased to $\sim$0.5 Crab in three weeks, to start
rising again to 1 Crab, and at the time of writing it is still brightening
(see Fig. 1).

We present here preliminary results
of the timing analysis of a subset of PCA observations: four
during the early phase of the outburst, when the source count rate was 
increasing monotonically with time, and one a few days into the 
decay. The times of the five observations 
considered here are marked with dotted lines in the bottom panel of Fig. 1.

\begin{figure}[htb]
\centering
\psfig{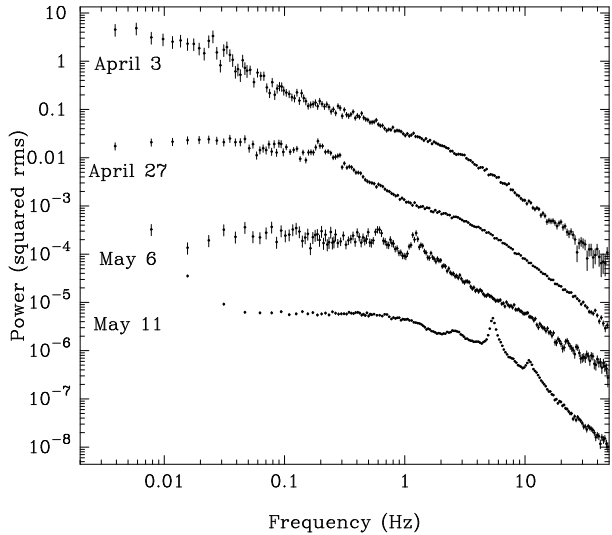}
\caption{Power Density Spectra of the first four observations marked in Fig.1.
The spectra have been shifted by factors of ten in squared rms for clarity.
}
\label{fig:ex}
\end{figure}

\section{Timing analysis of the early phase}

For each of the first four observations, we produced a Power Density Spectrum
(PDS) by selecting the energy range 3-15 keV, 
dividing the observation in shorter segments and averaging the corresponding
PDS. The resulting PDS can be seen in Figure 2. A strong evolution with time
is evident. The PDS of the first observation, on the initial rise of the
outburst, is very similar to the typical Low-State PDS observed before
with RTXE (e.g. \cite{b97,n02}). A weak QPO is seen at $\sim$0.25 Hz.
During the second observation, on the flattening part of the initial rise,
the PDS is similar, but shifted to higher frequencies (the QPO is now around
0.2 Hz). 
The third observation was made just before a sharp brightening. The
characteristic frequencies are even higher, with two harmonically-related
QPOs at 0.6 and 1.2 Hz. At a much higher count rate, the fourth PDS shows
even higher frequencies, with a 6 Hz QPO with complex harmonic content.
This PDS is almost indistinguishable from the VHS examples in \cite{miya}.
At the same time, the X-ray colors measured with the PCA show that the first
three observations follow a softening curve, while the fourth is much 
softer.

It is remarkable that we are observing here a smooth transition in the timing
properties from LS to VHS, with no discontinuity seen, while the energy
spectral parameters change abruptly.

\section{A transient QPO!}

Our fifth observation, on 2002 May 17, shortly after the ASM flux started
to decrease, is different. In the first RXTE orbit, the 1s binned light curve
does not show much variability and the PDS shows a power-law noise component
with a weak and broad bump between 6 and 10 Hz (Left panel in Fig. 3). 
In the second RXTE orbit, the power-law noise is stronger, as can be seen
from the light curve directly, and a very strong QPO at 6 Hz appears in the
PDS. A detailed analysis of a dynamical power density spectrum shows that 
this QPO is not present during the first 150 seconds of the orbit, then
appears suddenly within a time scale shorter than 16 seconds.

\begin{figure}[htb]
\centering
\psfig{file=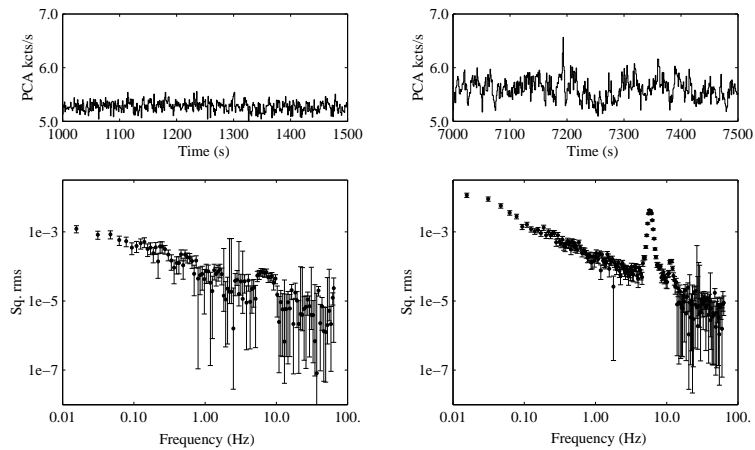,width=10cm}
\caption{Results for our fifth observation. 
500 s of light curve (bin 1 s) and PDS for the first (left panel) and 
second (right panel) RXTE orbit. 
}
\label{fig:ex}
\end{figure}

\section{A long-awaited outburst}
   
Our preliminary results show that the timing properties of the
source move gradually from the LS to the VHS, despite sudden changes in 
spectral shape. When at high flux, the source entered a HS, with reduced
timing variability, but the appearance of a strong transient QPO at 6 Hz
makes the HS interpretation questionable, as rapid oscillations between 
different ``flavors'' of VHS have already been observed \cite{miya}.



\begin{thebibliography}{}

\bibitem{atel}
 Smith D.M., et al., 2002, ATel., {\bf 95}.
\bibitem{mendez}
 M\'endez M., \& van der Klis M., 1997, ApJ, {\bf 479}, 926. 
 \bibitem{b97}
 Belloni T., et al., 1997, ApJ, {\bf 519}, L159.
 \bibitem{n02}
 Nowak M.A., Wilms J., \& Dove J.B., 2002, MNRAS, {\bf 332}, 856.
 \bibitem{miya}
 Miyamoto S., \& et al., 1991, ApJ, {\bf 383}, 784. 

\end{thebibliography}
\end{document}